\documentclass{article}

\usepackage{arxiv}

\usepackage[utf8]{inputenc} 
\usepackage[T1]{fontenc}    
\usepackage[hidelinks]{hyperref}       
\usepackage{url}            
\usepackage{booktabs}       
\usepackage{amsfonts}       
\usepackage{nicefrac}       
\usepackage{microtype}      
\usepackage[version=4]{mhchem}
\usepackage{url}
\usepackage{float}
\usepackage{doi}
\usepackage{subfigure}
\usepackage{multirow}
\usepackage{amsmath}
\usepackage{graphicx}
\usepackage{setspace}
\usepackage{amssymb} 
\usepackage{xcolor} 
\usepackage{subcaption}
\usepackage{tabularx}
\usepackage{xcolor,colortbl}

\usepackage[mathlines]{lineno}
\usepackage{appendix}
\usepackage{makecell} 

\definecolor{lavender}{rgb}{0.9, 0.9, 0.98}

\title{The traffic concentration effects of urban navigation services}

\date{}

\author{Giuliano Cornacchia \\
    ISTI-CNR, Pisa, Italy \\
    \texttt{giuliano.cornacchia@isti.cnr.it} \\
	\And
	Mirco Nanni \\
    ISTI-CNR, Pisa, Italy \\
    \texttt{mirco.nanni@isti.cnr.it}
   \And 
   Dino Pedreschi \\
    Department of Computer Science, Pisa, Italy \\
    \texttt{dino.pedreschi@unipi.it}
   \And 
   Luca Pappalardo\\
    ISTI-CNR, Pisa, Italy \\
    Scuola Normale Superiore, Pisa, Italy\\
    \texttt{luca.pappalardo@isti.cnr.it}
}

\begin{document}
\maketitle

\begin{abstract}
The collective impact of navigation services remains unclear: while often beneficial to individual drivers, they can unintentionally reshape urban traffic patterns. We simulate their impact in Florence, Milan, and Rome (Italy), integrating GPS data, road networks, and route recommendations from leading providers. We identify a concentration effect: as adoption increases, route diversity declines, and traffic and emissions converge onto fewer roads. At full adoption, route diversity decreases by up to 14\% compared to a baseline where recommendations are ignored. Moreover, navigation services reduce \ce{CO_2} emissions at low adoption levels, but these benefits diminish, disappear, or even reverse beyond a city- and service-specific threshold. We replicate our experiments in an abstract setting, obtaining results consistent with those observed in real-world cities.
\end{abstract}

\newpage
\section{Introduction}
\label{intro}

Navigation services (e.g., Google Maps and TomTom) recommend routes based on traffic conditions, guiding the movements of millions of drivers worldwide.
While beneficial to individual drivers, algorithmic navigation has been linked to numerous documented cases of localised congestion \cite{foderaro2017navigation, Siuhi2016-gb, hendrix2016traffic, schade2024traffic}. Yet, the trade-off between their benefits and detrimental effects on traffic dynamics remains poorly understood \cite{pappalardo2023future, pedreschi2025human}.

Research in this domain yields fragmented and contradictory results, primarily focused on specific navigation services and urban contexts \cite{pedreschi2025human, pappalardo2023future, pappalardo2025survey, cornacchia2022how}.
A simulation conducted by Google Maps in Salt Lake City reported measurable benefits from the use of its routing system, including an average reduction of 6.5\% in travel time and a 1.7\% decrease in \ce{CO_2} emissions \cite{arora2021quantifying}. 
Other studies find that some routing strategies can effectively mitigate \ce{CO_2} and \ce{NO_x} emissions \cite{barth2007environmental, ahn2013eco, valdes2016eco, perezprada2017managing}, reduce energy consumption \cite{barth2007environmental, ahn2013eco}, vehicle miles travelled \cite{thai2016negative}, and accident risks \cite{johnson2017beautiful}. 
Nevertheless, research also highlights unintended drawbacks, such as increased travel times, heightened exposure of populations to \ce{NO_x}, elevated traffic volumes in certain areas \cite{valdes2016eco, perezprada2017managing}, and a redistribution of traffic from highway to local streets, parks, and tourist destinations \cite{thai2016negative, johnson2017beautiful}.
These studies are difficult to generalise, replicate, and compare because they rely on specific data sources and methodologies. 
Overall, there is a lack of a robust, unified framework to assess the multifaceted impact of algorithmic navigation across multiple cities. 
This study quantifies the impact of various navigation services on road network usage and \ce{CO_2} emissions in three cities of different sizes.

We design a simulation framework that reproduces urban traffic dynamics by integrating mobility demand information from GPS data, road networks from OpenStreetMap, and route recommendations from several popular navigation services: Bing Maps, Google Maps, Mapbox, and TomTom. 
We use this framework to assess the impact of these navigation services in Florence, Milan, and Rome (Italy) by simulating different levels of service adoption. For each adoption rate, vehicles are randomly assigned to treatment and control groups, with the treatment group following algorithmic route recommendations.
This setup enables a causal assessment of how adoption rates of navigation services determine changes in traffic dynamics.

We find that route recommendations substantially reshape road network usage and \ce{CO_2} emissions.
Across all cities and navigation services, higher adoption rates cause a reduction in route diversity, i.e., recommended routes increasingly concentrate on a smaller set of roads. 
Under light traffic loads, navigation services reduce \ce{CO_2} emissions.
Under high traffic loads, navigation services reduce emissions at low adoption levels; however, these benefits weaken, vanish, or reverse as adoption increases.
To assess the generalisability of these findings, we replicate our experiments using an abstract representation of road networks and traffic dynamics, finding solid agreement with the patterns observed in real-world cities.

Our study sheds light on the multifaceted impact of algorithmic navigation across multiple cities and services, and it is replicable worldwide, given the widespread availability of mobility demand and road network data. 
Ultimately, our framework supports urban decision-makers in assessing the impacts of navigation services and designing targeted policy interventions. 

\section{Results}
\subsection{Simulation framework}
\label{sec:simulation_framework}
Our simulation framework leverages SUMO \cite{krajzewicz2002sumo, behrisch2011SUMOOverview, microscopic2018lopez}, an open-source, agent-based traffic simulator developed by the German Aerospace Center and used in several large-scale studies for its realism in replicating traffic dynamics \cite{argota2022getting, flitsch2018UpperAustriaSUMO, bachechi2020ModenaSensors, arora2021quantifying, cornacchia2022how}.
SUMO simulations capture the behaviour of a set of vehicles moving and interacting on a road network, including dynamics at road intersections, queues at traffic lights, and slowdowns due to traffic congestion.

A city’s road network is represented as a directed multigraph, where nodes correspond to road intersections and edges to road segments. We extract road networks from OpenStreetMap (see Methods for details). Urban mobility demand is modelled through an Origin-Destination (OD) matrix, which captures the volume of trips between different areas of the city. 
The OD matrix can be inferred from various data sources, including travel surveys, GPS data, and mobile phone data \cite{hazelton2001inference, cerqueira2022inference, pappalardo2022scikit, bonnel2018origin}.

We partition a city's area into equally sized square tiles of 1 km$^2$ and estimate OD flows between these tiles using GPS data describing trips of private vehicles (see Methods). Square tiles provide a spatially uniform discretisation, avoiding biases introduced by administrative boundaries (e.g., census zones or districts), which may vary in size, shape, and functional meaning. This choice ensures consistent spatial resolution across the city and facilitates comparability across urban areas.
The inferred OD matrix is then translated into a mobility demand, i.e., a set of individual vehicle trips that serve as inputs to the traffic simulation.
See Supplementary Note 1 for further details on GPS data and the computation of the mobility demand.

We assign routes to trips in the mobility demand using the public APIs provided by several navigation services.
These services recommend a route between an origin edge and a destination edge considering various factors, such as the typical traffic conditions at the time when the trip starts.
API requests allow us to simulate drivers' common queries, such as ``Take me from location A to location B by car at 8:30 AM''.
Although the algorithms underlying navigation services are opaque to us, we assume that they provide route recommendations optimised for individual requests rather than favouring system-level coordination. This is justified by their inability to enforce or forecast driver compliance, which makes system-optimal solutions based on Wardrop’s second principle \cite{wardrop1952proceedings, sheffi1984urban} unreachable.

We query routes for future dates relative to the request time, thereby forcing the APIs to rely on typical traffic information rather than real-time conditions. This choice avoids incorporating signals shaped by exogenous and unpredictable events -- such as accidents, strikes, or roadworks -- that cannot be modelled within the simulation.

\begin{figure}[!htb]
    \centering
    \includegraphics{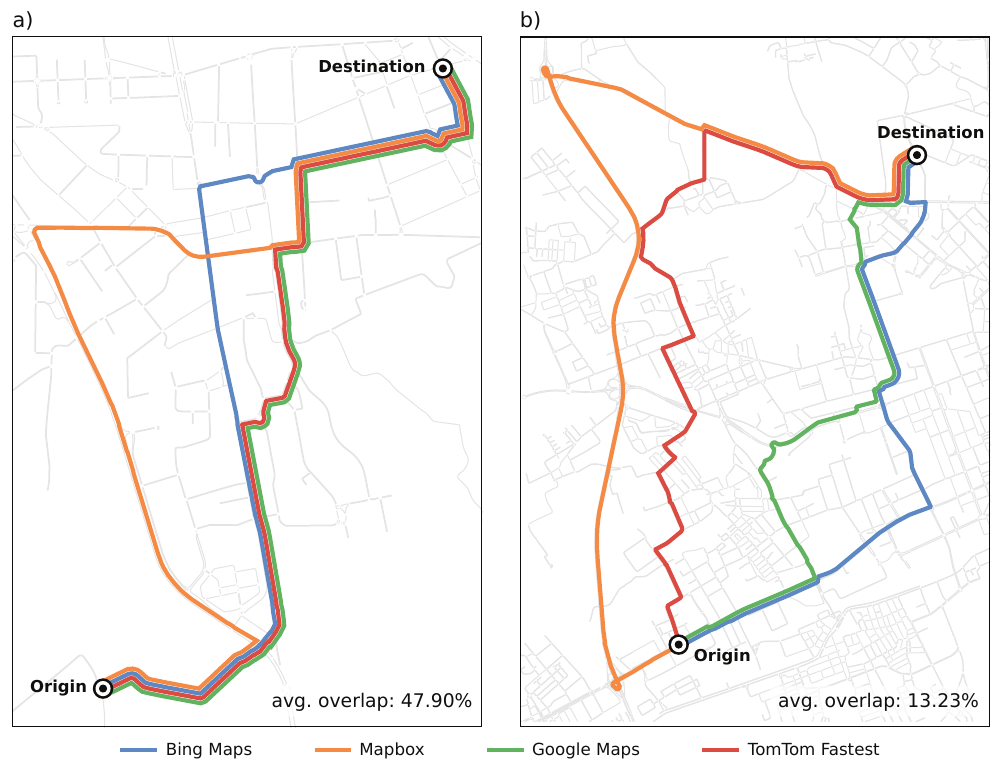}
    \caption{\textbf{Routes recommended by navigation services.} 
    (a, b) Two Origin-Destination pairs in Milan and the routes suggested by Bing Maps, Google Maps, Mapbox, and TomTom Fastest. (a) An illustrative example where recommendations largely overlap (average overlap = 47.90\%); (b) a case where recommendations largely diverge (average overlap = 13.23\%). The difference among navigation services occurs because they rely on different criteria and diverse historical data on traffic conditions. Road network data from OpenStreetMap (\textcopyright\ OpenStreetMap contributors), available under the Open Database License (ODbL; \url{https://www.openstreetmap.org/copyright}). OpenStreetMap documentation is published under CC BY-SA 2.0 (\url{https://creativecommons.org/licenses/by-sa/2.0/}).}
    \label{fig:figure1_od_trips}
\end{figure}

We consider a set of navigation services that provide public APIs: Bing Maps, Google Maps, Mapbox, TomTom Fastest, TomTom Shortest, and TomTom Eco-routing. At the time of the experiments, Apple Maps, Waze, and Google Maps eco-routing could not be included due to the unavailability of public APIs.
These services were selected because of their commercial relevance and widespread adoption, which makes them suitable for large-scale, reproducible experiments (see Supplementary Note 2 and Table \ref{tab:table_apis} for details). Different services may recommend different routes for the same trip (Fig.~\ref{fig:figure1_od_trips}a-b), reflecting differences in their optimisation objectives, heuristics, and underlying historical traffic data.
We find that the average route overlap across services ranges from 68\% to 97\% with an average of 76.4\% (see Supplementary Note 2).
We account for this variability by comparing results across multiple navigation services rather than relying on a single one. We refer to a trip that follows a route suggested by a navigation service $s$ as an $s$-routed trip.

\vspace{0.3cm}
\begingroup
\footnotesize
\centering
\begin{tabularx}{\linewidth}{c l l X}
\toprule
 & \textbf{provider} & \textbf{profile name} & \textbf{profile description} \\
\midrule
Bing Maps & Bing Maps & timeWithTraffic &
optimisation of travel time using current traffic information. \\
\rowcolor{lavender!80}
Google Maps & Google Maps & not available & not available \\
Mapbox & Mapbox & driving-traffic &
lowest probability of slowdowns given current and historical traffic conditions. \\
\rowcolor{lavender!80}
TomTom Eco-routing & TomTom & eco &
trade-off between travel time and fuel consumption. \\
TomTom Fastest & TomTom & fastest &
shortest travel time while keeping the routes sensible. \\ 
\rowcolor{lavender!80}
TomTom Shortest & TomTom & short &
trade-off between travel time and travel distance. \\
\bottomrule
\end{tabularx}
\captionof{table}{\textbf{Characteristics of selected navigation services.}
The table shows the service provider, the name of the profile used
(i.e., the routing criteria) and a description of the criteria.}
\label{tab:table_apis}
\endgroup
\vspace{0.3cm}

To assess the impact of a navigation service $s$
on traffic in a given city, we conduct controlled experiments by varying the adoption rate of $s$ from $r{=}0\%$ (no adoption) to $r{=}100\%$ (full adoption) in steps of 10\%.
We assign $r\%$ of vehicles to a treatment group and the remaining vehicles to a control group. 
Vehicles in the treatment group are $s$-routed, while those in the control group choose a route autonomously based on an approximate knowledge of the road network and traffic patterns.
To emulate this approximate knowledge, control vehicles follow a modified version of the fastest route computed by SUMO.
This modification introduces a small random deviation from the fastest route to account for imperfections in human route choice, due to limited network knowledge, bounded rationality, and cognitive biases documented in the transportation and behavioural literature \cite{seele2012cognitive, di2016boundedly, zhu2015do}. The full implementation details of this modification are described in Methods.
The similarity between SUMO's fastest routes and those suggested by navigation services lies in the 52-63\% range with an average of 57.2\%. 
This overlap is substantially lower than the similarity observed among navigation services themselves (68-97\%), indicating that SUMO relies on different routing criteria and cost functions.

For statistical robustness, we repeat the simulation for a given adoption rate $r$ ten times, using different random compositions of the treatment and control groups in each run. 
Furthermore, we repeat all the above steps considering both low and high traffic loads.
Low traffic loads refer to situations with few circulating vehicles in the city, such as off-peak hours and nighttime.
High traffic load indicates that the traffic is approaching congestion.
See Supplementary Note 3 for details on the definition of low and high traffic loads.
 
We evaluate the impact of navigation services both in isolation, where all the vehicles in the simulation use the same service, and in combination, where vehicles may choose among different navigation services.
We present only the isolation scenario since the results are analogous (see Supplementary Note 2 for results on the combination scenarios).
In our analysis, we focus on two measures: route diversity and \ce{CO_2} emissions. 
Route diversity captures the spatial concentration of routes across the road network and it is defined as the number of edges traversed at least once by any vehicle. 
\ce{CO_2} emissions provide an aggregate, environmentally relevant outcome of traffic dynamics, integrating the effects of congestion, speed fluctuations, and acceleration patterns. 
We compute this measure using the microscopic emission model implemented in SUMO \cite{microscopic2018lopez}, which estimates instantaneous vehicle emissions as a function of speed and acceleration. See Methods for details on the implementation of the two metrics.

A limitation of our experimental setup arises from interaction effects: since vehicles move on a shared road network, those assigned to the control group inevitably interact with vehicles in the treatment group \cite{cox1958planning, aronow2017estimating}. 
As a result, the outcomes of individual vehicles are not fully independent and observable differences between groups are expected to be underestimated. 
This is a common situation when performing controlled experiments in complex social systems \cite{huszar2022algorithmic}.
Our results should therefore be interpreted as conservative estimates
of the impact induced by navigation services.

\subsection{Impact of navigation services}
\label{sec:impact}

We conduct experiments in three cities in Italy: Florence, Milan, and Rome. 
These cities were chosen due to their heterogeneity in size, population, and road network structure (See Fig. S1). 
For instance, Rome has an extensive and sparse road network with relatively long edges and moderately winding roads. 
Milan is slightly smaller and denser than Rome and is characterised by shorter edges and a more regular network structure. 
Florence, the smallest, has the most winding and spatially intricate network. The difference among cities enables assessing the robustness of algorithmic navigation effects across different urban morphologies.
We simulate traffic in each city during the morning peak, defined from vehicular GPS data as the one-hour morning slot with the highest traffic intensity.

\begin{figure}[H]
    \centering
    \includegraphics{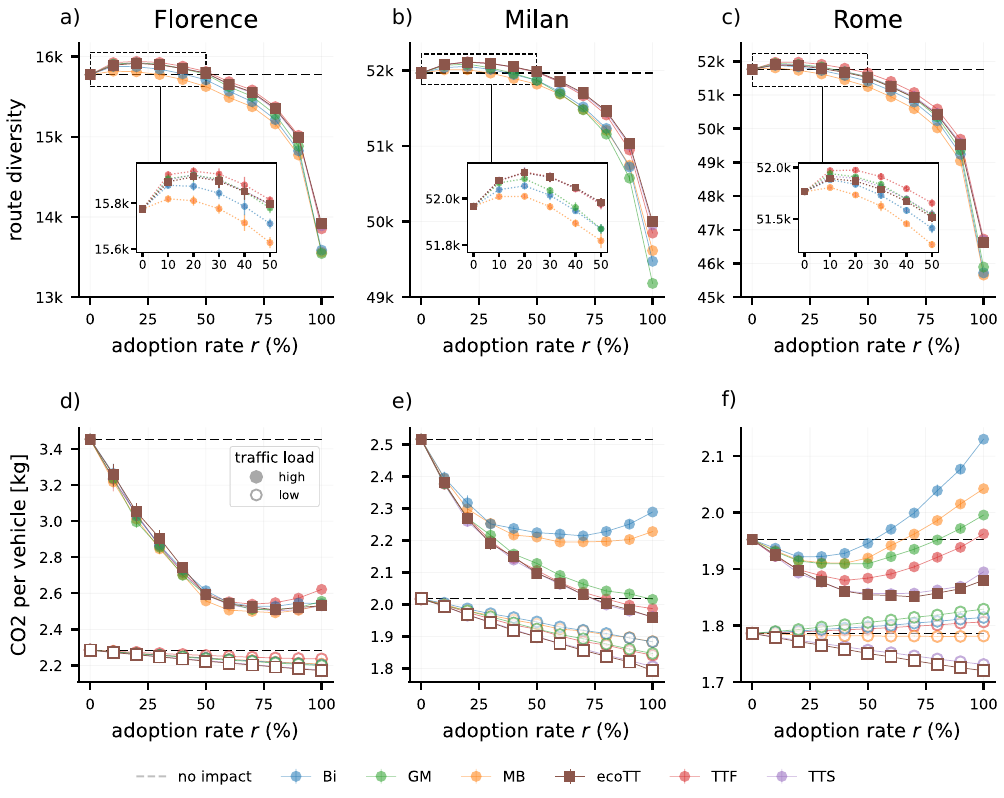}
    \caption{\textbf{Impact of navigation services on route diversity and \ce{CO_2} emissions.}
    (a-c) Service adoption rate, $r$, versus route diversity at high traffic loads in Florence, Milan, and Rome (Italy).  
    The inset plots zoom on the range $r{=}0\%, \dots, 50\%$, to highlight the slight initial increase of route diversity. 
    Increased adoption of navigation services causes a reduction in route diversity, with only minor fluctuations among simulation runs, cities, and navigation services.
    (d-f) Service adoption rate versus average \ce{CO_2} emissions per vehicle at high traffic loads (filled markers) and low traffic loads (empty markers).  
    At high traffic loads, when $r$ is low, \ce{CO_2} emissions decrease considerably; when $r$ exceeds a certain city- and service-dependent threshold, the benefits plateau and, in some cases, even reverse.
    In all panels a-f, the dashed line represents the $r{=}0\%$ scenario; markers indicate the average route diversity or the average \ce{CO_2} emissions over ten simulations with different random choices of $s$-routed vehicles. Square markers refer to the navigation service ecoTT, which employs eco-routing. 
    Vertical bars, indicating the standard deviation across the ten simulations, are hardly visible because they are negligible. Navigation services are indicated as follows: Bi (Bing Maps), GM (Google Maps), MB (Mapbox), TTF (TomTom Fastest), TTS (TomTom Shortest), and ecoTT (TomTom Eco-routing).}
    \label{fig:figure2_route_div_co2}
\end{figure}

Fig. \ref{fig:figure2_route_div_co2}a-c illustrates the impact of varying the adoption rate $r$ on route diversity in the three cities, considering high traffic load.
Results under low traffic load are similar (see Supplementary Fig. S2). 
If the navigation service had a negligible impact, we would expect a constant route diversity as $r$ increases (horizontal dashed line in Fig. \ref{fig:figure2_route_div_co2}a-c). 
Instead, route diversity varies with $r$ following an exponential function (see Supplementary Note 4). 
When $r$ is below a specific city- and service-dependent threshold (varying from about $25\%$ to $50\%$), route diversity slightly increases by 0.15-1.05\% in Florence, 0.08-0.34\% in Milan, and 0.01-0.41\% in Rome (see inset plots in Fig. \ref{fig:figure2_route_div_co2}a-c) compared to $r{=}0\%$.
On the other hand, when $r$ exceeds this threshold, route diversity considerably decreases.
At the full adoption rate ($r{=}100\%$), route diversity is reduced by 11.80-14.34\% in Florence, 3.79-6.87\% in Milan, and 9.73-14.26\% in Rome compared to $r{=}0\%$.
We also examine the distribution of route diversity separately for vehicles in the two groups: vehicles in the treatment group exhibit lower route diversity compared to those in the control group, regardless of the adoption rate (see Supplementary Note 4). 
Our results have only minor fluctuations among different random compositions of the treatment and control groups and across cities and navigation services. 
Therefore, changes in route diversity can be confidently attributed to the adoption of algorithmic navigation.

Fig. \ref{fig:figure2_route_div_co2}d-f shows the average \ce{CO_2} emissions per vehicle produced at different traffic loads and adoption rates.
When the traffic load is low, navigation services are generally beneficial: in most cases, \ce{CO_2} emissions are lower than in the $r{=}0\%$ scenario, showing a near-linear decreasing trend with $r$. 
At the full adoption rate, navigation services reduce \ce{CO_2} emissions by 2.15-5\% in Florence and 6.64-11.13\% in Milan. 
In Rome, some navigation services reduce \ce{CO_2} emissions by 0.26-3.69\%, while Google Maps, TomTom Fastest and Bing Maps slightly increase them by 2.4\%, 1.13\% and 1.57\%, respectively. See Table S2 for a detailed comparison across cities and navigation services.

However, at high traffic loads, the impact of navigation services depends on the adoption rate. 
When $r$ is low, \ce{CO_2} emissions decrease; when $r$ exceeds a certain city- and service-dependent threshold, the benefits plateau and in some cases \ce{CO_2} emissions increase (see Fig. \ref{fig:figure2_route_div_co2}d-f). 
At full adoption ($r{=}100\%$), reduced route diversity leads to a concentration of \ce{CO_2} emissions on fewer roads, thereby increasing inequality in their spatial distribution (see Supplementary Note 5).
Fig. \ref{fig:figure3_maps} illustrates this effect for TomTom Fastest: at $r{=}0\%$ (Fig, \ref{fig:figure3_maps}a, d, g), the traffic is more evenly distributed, whereas at $r{=}100\%$ (Fig. \ref{fig:figure3_maps}b, e, h), there is a concentration of traffic and \ce{CO_2} emissions on fewer edges.
This effect is further illustrated by the Lorenz curves and corresponding Gini coefficients of \ce{CO_2} emissions across road segments (Supplementary Fig. S10), reported for adoption rates $r{=}0\%$, $r{=}50\%$, and $r{=}100\%$.
In Florence and Milan, emissions are more evenly distributed at $r{=}50\%$ than at either 0\% or 100\%, as indicated by a lower Gini coefficient, reflecting a temporary increase in route diversity before route convergence takes hold.

At $r=100\%$, emissions become more concentrated: the Gini coefficient increases by 0.023 in Florence, 0.025 in Milan, and 0.031 in Rome compared to 50\% adoption. This corresponds to an additional 2.04\% (Florence), 2.31\% (Milan), and 3.9\% (Rome) of total \ce{CO_2} emissions being concentrated on the top 20\% most polluted edges.

Across all cities, we observe a consistent shift of routes, and consequently \ce{CO_2} emissions, toward high-capacity road infrastructures such as highways and major arterial corridors (see Fig. \ref{fig:figure3_maps}c, f, i).
These infrastructures are typically favoured by navigation services because they offer higher speed limits and accommodate large traffic volumes.
However, despite accounting for only about 6\% of the road network, these high-capacity road infrastructures absorb a large share of \ce{CO_2} emissions. This imbalance becomes even more pronounced as navigation service adoption increases. From 0\% to 100\% adoption, the share of emissions on high-capacity infrastructures rises from 17.61\% to 36.25\% in Florence, 9.94\% to 26.42\% in Milan, and 26.02\% to 33.66\% in Rome.
The full breakdown of \ce{CO_2} emissions by road type is provided in Supplementary Note 5.

\begin{figure}
    \centering
    \includegraphics[width=0.9\linewidth]{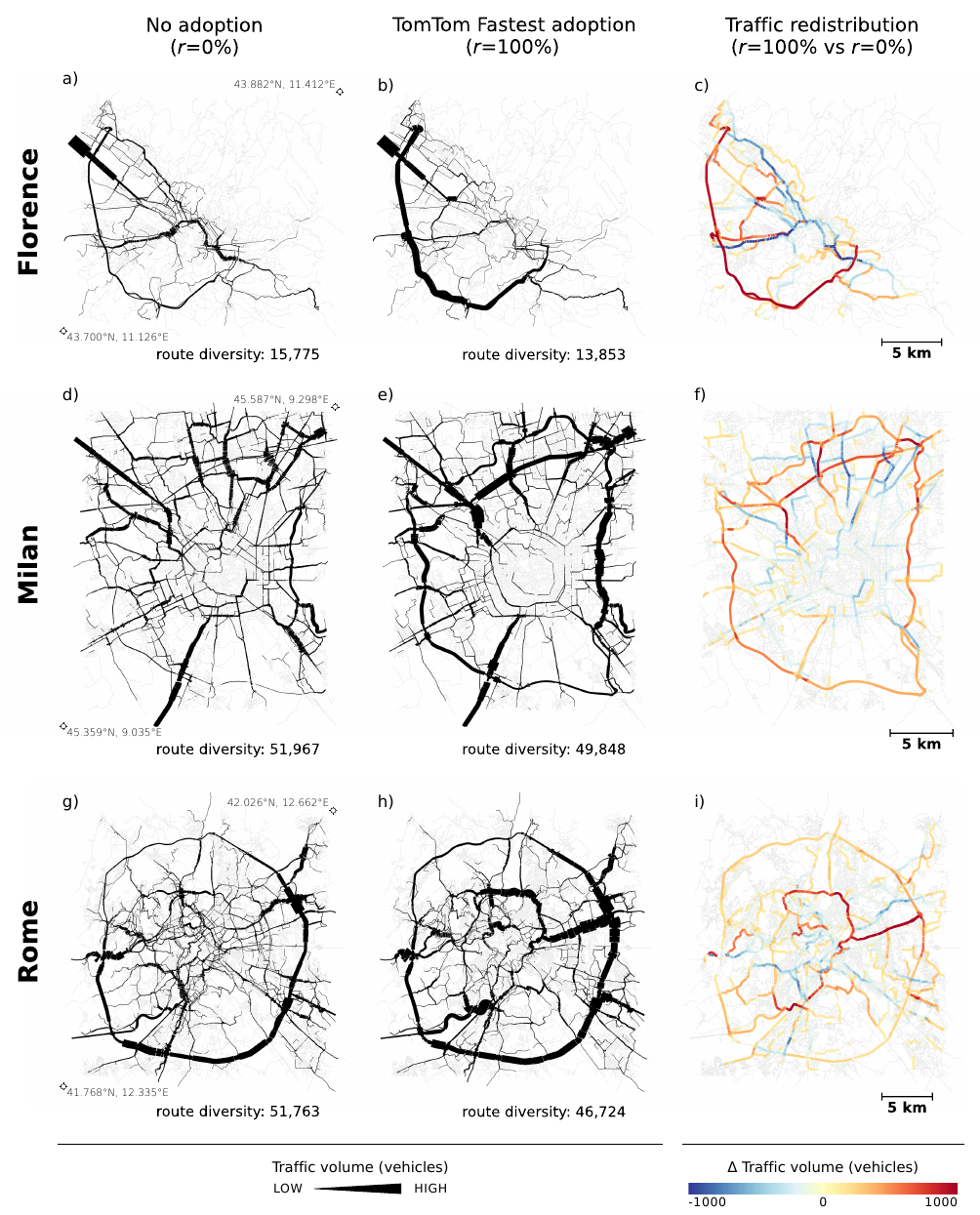}
    \caption{\textbf{Redistribution of routes.}
    Rows correspond to cities: Florence (a-c), Milan (d-f), and Rome (g-i). 
    Columns represent navigation scenarios under TomTom Fastest (TTF): no navigation adoption (a, d, g), full adoption (b, e, h), and traffic redistribution (c, f, i). 
    For the no navigation and full navigation scenarios, road edge widths are proportional to traffic volume. 
    The redistribution scenario shows differences in road usage between the full adoption and no adoption scenarios: blue edges denote road segments where the full adoption of TTF reduces traffic relative to the no adoption scenario; red edges indicate segments where traffic increases. Colour intensity is proportional to the magnitude of the traffic change. 
    Across all cities, full TTF adoption concentrates traffic onto a smaller subset of road segments, with traffic increasingly redistributed onto high-capacity roads. Road network data from OpenStreetMap (\textcopyright\ OpenStreetMap contributors), available under the Open Database License (ODbL; \url{https://www.openstreetmap.org/copyright}). OpenStreetMap documentation is published under CC BY-SA 2.0 (\url{https://creativecommons.org/licenses/by-sa/2.0/}).}
    \label{fig:figure3_maps}
\end{figure}

Results further indicate city-specific variations in the observed effects.
In Florence, the largest reduction in average \ce{CO_2} emissions occurs at $r{=}70$-$80\%$; beyond this point, the benefits diminish (see Fig. \ref{fig:figure2_route_div_co2}d).
In Milan, Google Maps and TomTom-based services reduce \ce{CO_2} emissions as $r$ increases, achieving reductions of 19-22\%  at the full adoption rate.
In contrast, Mapbox and Bing Maps reach their highest reductions ($\approx$12\%) when $r{=}60\%$ and $r{=}70\%$, respectively; beyond these rates, the benefits diminish (\ce{CO_2} reductions decrease by 9\% compared to $r{=}0\%$). 
In Rome, optimal adoption rates vary by service: 20\% for Bing Maps, 40\% for Mapbox, Google Maps and TomTom Fastest, 60\% for TomTom Shortest, and 70\% for TomTom Eco-routing, leading to an average \ce{CO_2} reduction of 3.29\%. 
Notably, Bing Maps, Mapbox, Google Maps and TomTom Fastest lead to increased \ce{CO_2} emissions beyond certain thresholds compared to $r{=}0\%$: 9.10\% for Bing Maps, 4.61\% for Mapbox, 2.22\% for Google Maps and 0.52\% for TomTom Fastest (see Supplementary Table S2 for details on the \ce{CO_2} reductions). In general, we observe an adoption range for most navigation services in the three cities -- between 60\% and 80\% -- beyond which the environmental benefits of algorithmic navigation vanish.

A comparison of Florence, Milan, and Rome reveals both common and city-specific patterns: route concentration is a consistent outcome of increased service adoption, while \ce{CO_2} emissions vary with the spatial redistribution of traffic induced by navigation services.
In Florence, navigation services shift traffic toward an arterial road that is moderately used in the no-adoption scenario (Fig. \ref{fig:figure3_maps}a-b). This redistribution is associated with a substantial reduction in CO$_2$ emissions that stabilises as this corridor becomes saturated at high adoption levels.
In Rome, traffic is redirected onto a few major roads that are already heavily used in the baseline scenario (Fig. \ref{fig:figure3_maps}g-h). This pattern may explain the pronounced increase in \ce{CO_2} emissions at high adoption rates. 
Milan lies between these two cases, navigation services concentrate traffic onto several major roads that are lightly trafficked in the baseline scenario (Fig. \ref{fig:figure3_maps}d-e), which is consistent with the moderate decreases in \ce{CO_2} emissions observed at high adoption rates.

Our results show that even an eco-routing service (TomTom Eco-routing) exhibits \ce{CO_2} trends comparable to those of other navigation services (see brown squares in Fig.~\ref{fig:figure2_route_div_co2}). This indicates that optimising routes at the individual level with explicit environmental objectives does not necessarily translate into reduced emissions at the system level. Environmental sustainability depends not only on the properties of individual routes, but also on the collective distribution of routing decisions across the network.

To further investigate the relationship between route diversity and \ce{CO_2} emissions under high traffic load, we examine the correlation between the marginal changes in route diversity ($\Delta D_r$) and those in \ce{CO_2} emissions ($\Delta E_r$). 
$\Delta D_r$ is defined as the difference in route diversity between adoption rates $r{-}10\%$ and $r$, and $\Delta E_r$ as the difference in total \ce{CO_2} emissions over the same interval. 
Consequently, positive values of $\Delta D_r$ and $\Delta E_r$ indicate reductions in route diversity and \ce{CO_2} emissions, respectively.

\begin{figure}
    \centering
    \includegraphics{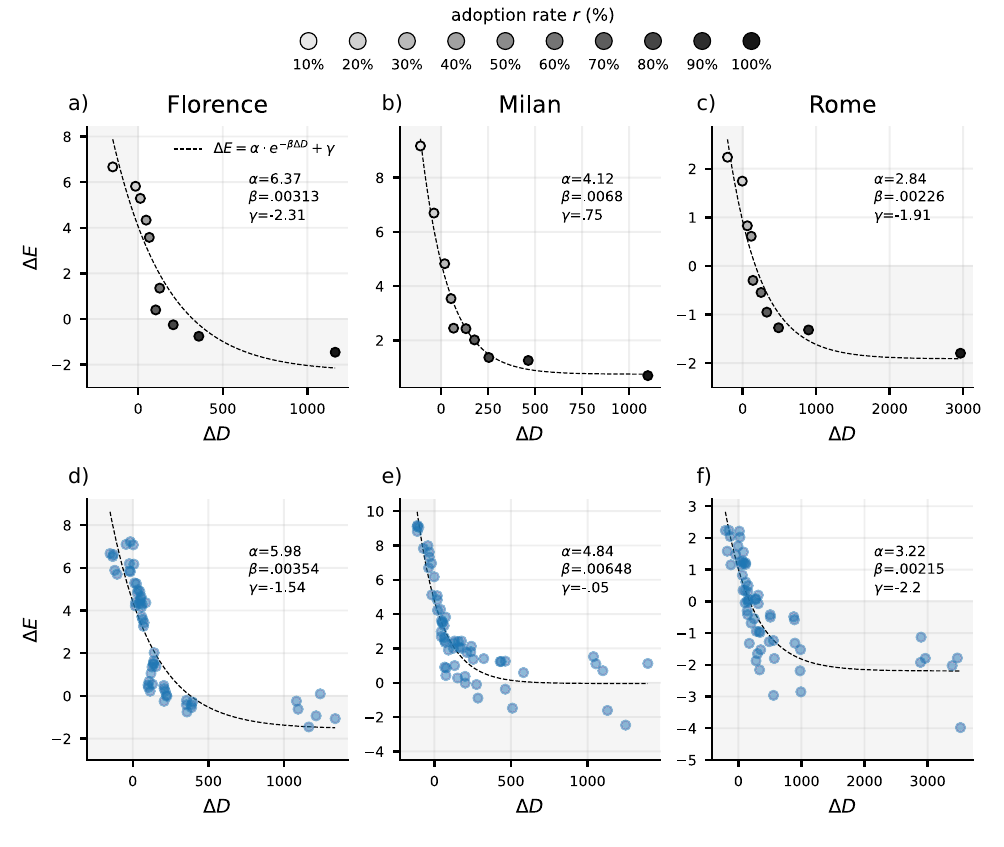}
    \caption{\textbf{Relationship between route diversity and \ce{CO_2} emissions.} 
(a-c) The relationship between the marginal change in route diversity ($\Delta D$) and the marginal change in \ce{CO_2} emissions ($\Delta E$) for TomTom Fastest (TTF) in Florence, Milan, and Rome (Italy). We find similar results for the other navigation services. The points are shaded from light grey to black, representing the service adoption rate (from $r{=}10\%$ to $r{=}100\%$).  
(d-f) Same relationship for all navigation services. Regions where $\Delta D$ or $\Delta E$ are negative are highlighted in grey. 
The black dashed line represents the exponential decay fit for each scenario.
At low adoption rates, slight increases in route diversity lead to substantial reductions in \ce{CO_2} emissions. As $r$ increases, small reductions in route diversity result in moderate \ce{CO_2} reductions. However, as $\Delta D$ further increases, $\Delta E$ decreases, indicating a diminishing return effect. This pattern is consistent across all cities and navigation services.}
\label{fig:co2_vs_concentration}
\end{figure}

Across all cities, we observe a strong negative association between $\Delta D_r$ and $\Delta E_r$: Spearman’s rank correlation coefficient is $\rho {=} -0.958$ in Florence, $\rho {=} -0.882$ in Milan, and $\rho {=} -0.899$ in Rome.  This association is well described by an exponential decay function (see Fig. \ref{fig:co2_vs_concentration}a-c).
At low adoption rates, small increases in route diversity are associated with large reductions in \ce{CO_2} emissions.
As the adoption rate increases, small reductions in route diversity result in moderate \ce{CO_2} reductions. Further losses in route diversity are accompanied by little or no additional benefit, indicating a diminishing return effect.
This pattern is consistent across all cities and navigation services (Fig. \ref{fig:co2_vs_concentration}d-f).

\subsection{Driving factors of concentration patterns}

To demonstrate that the observed concentration effects do not depend merely on network layouts and mobility demands of the three cities, we replicate our experiments in an abstract setting.
We model the road network as a regular grid in which all roads are bidirectional and have identical speed limits. This grid structure enables precise control of network topology while preserving essential flow interaction properties.
As the grid network is synthetic, we cannot rely on navigation service APIs. Instead, we emulate navigation services by assigning vehicles the fastest route, which provides an individually optimised route to the destination. Accordingly, $s$-routed vehicles follow the fastest path, while the remaining vehicles are assigned perturbed alternatives, consistent with our experiments on real-world cities.

\begin{figure}
    \centering
    \includegraphics{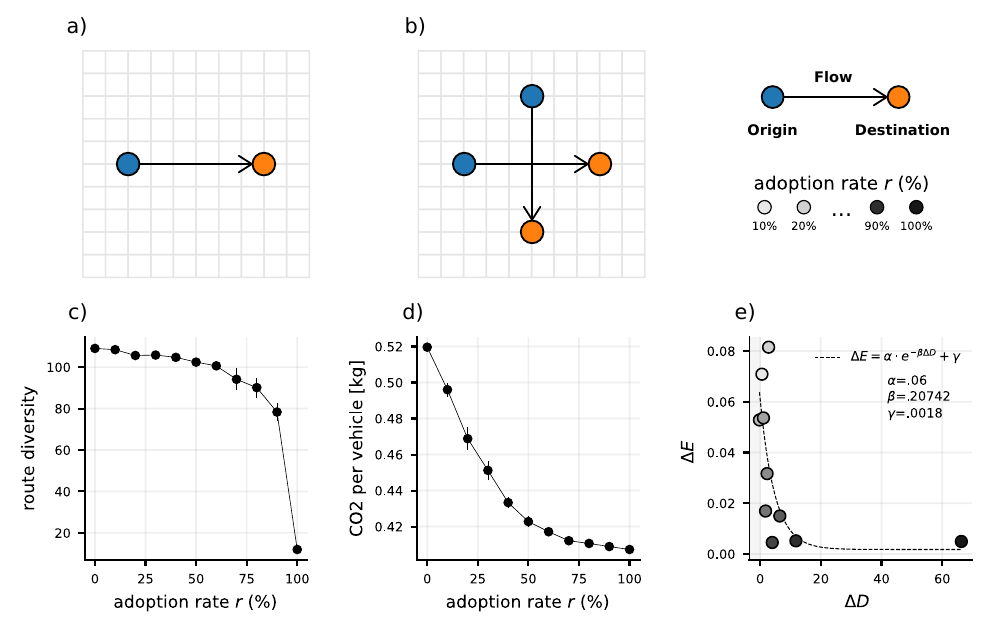}
    \caption{\textbf{Impact of algorithmic routing in an abstract setting.} 
(a, b) A $10{\times}10$ grid network with synthetic Origin-Destination (OD) flows. 
(a) Setting with a single OD flow. (b) Setting with two intersecting OD flows that increase vehicle interactions at the grid centre. 
Routed vehicles follow the fastest route, while non-routed vehicles take perturbed variants. (c) Route diversity as a function of adoption rate $r$. With two OD flows, route diversity decreases rapidly as $r$ increases, replicating patterns observed in real-world cities. (d) Average \ce{CO_2} emissions per vehicle versus adoption rate. Emissions initially decline, then plateau due to increased congestion at shared intersections. (e) Marginal relationship between changes in route diversity ($\Delta D$) and \ce{CO_2} emissions ($\Delta E$), showing a nonlinear trend consistent with real-world cities. Points are shaded from light grey to black to represent increasing adoption rates from $r{=}10\%$ to $r{=}100\%$.}
\label{fig:results_simplified_model}
\end{figure}

We first simulate a single OD flow on a $10 {\times} 10$ grid network (see Fig. \ref{fig:results_simplified_model}a), where all vehicles share the same origin and destination. We vary the adoption rate as in the experiments on real-world cities.
We find that, as the adoption rate increases, route diversity declines but \ce{CO_2} emissions continue to decrease (see Supplementary Fig. S13). This suggests that, in the absence of vehicle interactions at intersections, reduced route diversity alone does not necessarily lead to a diminishing return effect.

To enforce vehicle interactions, we simulate two orthogonal OD flows on the same grid (see Fig. \ref{fig:results_simplified_model}b). In this configuration, the fastest routes intersect at the centre of the grid, becoming interdependent and producing collective concentration effects.
All key patterns observed in real-world cities as the adoption rate increases are reproduced (see Fig. \ref{fig:results_simplified_model}c-e): a sharp reduction in route diversity, a plateau in \ce{CO_2} emissions, and a nonlinear relationship between $\Delta D_r$ and $\Delta E_r$. 
The simple abstract setting also captures the rise in \ce{CO_2} inequality as adoption rate approaches 100\%, with the increasing Gini coefficient aligning closely with the patterns observed in real-world cities (see Supplementary Fig. S19).

We extend the analysis to a larger grid and randomised grid-like networks and find that the same patterns persist (see Supplementary Note 6 for full details). This confirms that the observed concentration patterns are not artefacts of network layouts.

\section{Discussion}
\label{discussion}

In all three cities, as well as in the abstract setting, we observe a consistent pattern: increasing the adoption rate of navigation services leads to an increased concentration of routes and \ce{CO_2} emissions.
These results are consistent with findings from other domains, where algorithms and AI systems have been shown to reinforce conformist behaviour and existing inequalities (e.g., in the recommendation of urban venues \cite{mauro2026urban, sanchez2023bias} and products \cite{lee2019recommender} as well as in content generation \cite{shumailov2024ai}).

We note that the urban road networks considered in this study are not unconstrained systems. Italian cities are characterised by the presence of regulated traffic zones, such as limited traffic zones (ZTLs), and, in the case of Milan, a low-emission zone (LEZ), which restrict access to specific areas or vehicle types. These regulatory constraints are already incorporated into the routing logic of navigation services and therefore into our simulations. 
The concentration effect we observe arises from the alignment of route recommendations across drivers and is expected to generalise to other urban contexts, even though the specific roads involved may differ.

The spatial concentration observed in our study emerges as a collective property of algorithmic route recommendations. 
Based on our results, we argue that this effect stems from the tendency of navigation services to suggest similar routes in response to similar requests. Optimal routes are typically funnelled through the same efficient roads; as a consequence, many optimal routes overlap on a limited set of road edges, producing traffic concentration on such edges. The provision of a small number of alternative routes per request (typically no more than three) can only marginally dilute this concentration and cannot eliminate it. Moreover, the coexistence of multiple navigation services is unlikely to mitigate the effect, as competing platforms rely on comparable optimisation principles and therefore produce aligned routing recommendations. The observed concentration pattern carries important implications for urban inequality, environmental sustainability, urban policy, and the governance of digital platforms.

The first implication concerns urban inequality. At high adoption rates, navigation services tend to funnel traffic onto highways and major arterial roads. 
This spatial redistribution carries various consequences.
First, neighbourhoods located near highways and major arterial roads -- often situated at the urban periphery -- are exposed to higher levels of emissions, pollution, and noise, potentially leading to a decline in quality of life and property values.
Consequently, the observed concentration effect may exacerbate existing social disparities, in line with previous studies showing how algorithmic decision-making can amplify inequalities in urban environments \cite{johnson2017beautiful,ensign2018runaway, harvey1996justice, schlosberg2007defining}.

Another key implication concerns environmental sustainability: the impact of navigation services is non-negligible and depends on both traffic load and adoption rates. While algorithmic navigation tends to reduce emissions under low-traffic conditions and low adoption, its effects become mixed at a higher traffic load, with benefits diminishing, vanishing, or even reversing as adoption increases. This context dependence helps explain the fragmentation and apparent contradictions in the existing literature, which often focuses on specific traffic conditions or adoption levels, and shows that the impact of algorithmic navigation cannot be reduced to a simple dichotomy of beneficial or detrimental effects.
Notably, TomTom Eco-routing follows the same qualitative dynamics as other navigation services, highlighting a limitation of individualised optimisation to address the environmental footprint of algorithmic navigation.
When many drivers are guided toward similar “eco-optimal” paths, traffic concentration on these paths can offset or even reverse the intended environmental benefits.

Third, our simulation framework offers a valuable tool for urban policymaking, as it can be readily instantiated across cities worldwide. 
The framework allows simulating traffic outcomes across different adoption levels, including counterfactual scenarios in which navigation services are not used. 
This counterfactual analysis may help assess whether the impacts of algorithmic navigation align with or undermine existing or prospective municipal policies.
For example, by comparing the actual traffic with a baseline scenario in which navigation services are not used, policymakers can identify how algorithmic navigation reshapes urban mobility, e.g., amplifying traffic in sensitive areas such as school and hospital districts. 
This knowledge can inform targeted requests to navigation service providers aimed at mitigating or compensating for such effects.
In this sense, our framework aligns with current regulations -- such as the EU’s Digital Services Act \cite{EU2022DSA} -- which requires very large online platforms to detect and understand the systemic risks they pose to society in terms of fundamental rights, public security, and well-being.

Our study provides a systematic investigation of the impact of algorithmic navigation on urban spaces across multiple cities, navigation services, traffic loads, and adoption rates.
Our framework allows the simulation of hypothetical yet realistic scenarios that are difficult, if not impossible, to investigate empirically.
Direct on-road experiments face several practical and methodological constraints.
First, even if a navigation service provider conducted a controlled empirical study, it would capture only a subset of the urban vehicle fleet, leaving non-user movements and interactions unobserved.
Second, these controlled studies are difficult to replicate under identical initial conditions, as traffic dynamics are continuously shaped by unpredictable exogenous factors, including mobility demand fluctuations, car accidents, and roadworks. Third, large-scale experiments would require explicit driver consent and coordinated interventions at the urban level, raising critical logistical, regulatory, and ethical challenges.
Although simulations can provide only estimates rather than exact measurements, they currently represent a feasible approach for systematically studying the impact of navigation services at scale.

Our simulation framework can also be extended to study other urban mobility externalities associated with navigation services, including traffic safety (e.g., the risk and spatial distribution of car accidents) and neighbourhood-level segregation effects (e.g., the systematic avoidance of certain areas).
Beyond navigation services, our simulation framework can be extended to other mobility services, such as car sharing and ride hailing. These services exploit matching and dispatching algorithms that associate drivers with users to serve rides on demand \cite{santi2014quantifying}.
Recent research shows that these services have both beneficial and detrimental effects on traffic congestion, emissions, public transport usage, and social equity \cite{pappalardo2025survey, ngo2021effects, yan2020fairness}.
We can adapt our framework to simulate the impact of these services on various urban externalities.
For instance, driver-user matching can be implemented using standard strategies such as assigning the closest idle vehicle \cite{bokanyi2020understanding}, while rides can be assigned to routes either through conventional routing algorithms or by querying navigation service APIs. As in our experimental setup, the adoption rate of these services can then be systematically varied to quantify their impact on route diversity, traffic distribution, and different forms of emissions.

Finally, our study contributes to the broader discussion on human-AI coevolution \cite{pedreschi2025human}. Navigation services operate within a dynamic feedback loop: traffic patterns shape algorithmic recommendations, these recommendations influence individual route choices, and the resulting collective behaviour modifies future traffic conditions, in a potentially never-ending cycle \cite{pedreschi2025human, rahwan2019machine}. 
Recent work shows that feedback loop mechanisms can amplify existing biases in human-AI ecosystems, spanning domains from social media and online retail to human mobility \cite{pappalardo2025survey}. 
In this context, the interdependence between adoption rates, route diversity, and \ce{CO_2} emissions suggests that the concentration effects observed in our non-dynamic setting may be exacerbated when navigation services and users continuously adapt to one another within a feedback loop. 
Therefore, our results are probably an underestimation of the real impact, and a natural direction for future research is to investigate how the observed impact of navigation services evolves in a coevolutionary setting.

Looking ahead, the key challenge is to design navigation services that preserve route diversity, improving individual travel without inducing collective concentration. While our framework can be used to evaluate the impacts of alternative routing strategies, developing algorithms that account for the city as a complex system remains an open research question.

\section{Methods}
\label{methods}

\subsection{Road Networks}

We model a city’s road network as a directed multigraph, where nodes correspond to road intersections and edges to road segments.
We extract the road networks of Florence, Milan, and Rome (Italy) from OpenStreetMap (OSM) using SUMO's OSM Web Wizard. 
We manually fine-tune the road networks to correct inaccuracies that could 
potentially cause deadlocks and other unrealistic behaviours during simulations.
This fine-tuning phase includes correcting lane number inaccuracies, addressing 
road continuity disruptions, and modifying turns to align with real-world 
conditions~\cite{argota2022getting}. 
We use Google Maps and Google Street View as benchmarks to ensure the accuracy of these adjustments. Overall, we manually corrected $< 0.15\%$ of the road edges in the network.

\subsection{Impact measures}
We measure the impact of navigation services on route diversity and \ce{CO_2} emissions.
Route diversity captures the extent to which a set of routes explores the road network.
Formally, let $R$ be a set of routes, route diversity is defined as $\mathrm{RD}(R) = |\bigcup_{r \in R} \{ e \in r\}| $, where $e \in r$ indicates an edge in route $r$.
Higher values of route diversity indicate a broader exploration of the network, whereas lower values reflect convergence onto a limited subset of edges. 

Vehicle \ce{CO_2} emissions are computed using the emission modelling framework implemented in SUMO, based on the HBEFA3/PC\_G\_EU4 model \cite{infras2013handbook}. This model estimates instantaneous emissions at each trajectory point $j$ as $\mathcal{E}(j) = c_0 + c_1sa + c_2sa^2 + c_3s + c_4s^2 + c_5s^3$, where $s$ and $a$ are the vehicle's speed and acceleration at point $j$, respectively, and $c_0,\dots,c_5$ are vehicle- and emission-specific parameters extracted from the HBEFA database \cite{krajzewicz2015second}, corresponding to SUMO’s default vehicle emission class, i.e., a gasoline-powered Euro 4 passenger car.

We find that \ce{CO_2} emissions are strongly correlated with other measures, including fuel consumption, \ce{NO_x} emissions, particulate matter, and noise, at both the road-segment and per-vehicle levels (Pearson correlation $r>0.994$ across all cities and navigation services). We therefore report results for \ce{CO_2} emissions only.

\subsection{Modified fastest route}
Vehicles in the control group follow a modified version of the fastest route implemented by SUMO's duarouter algorithm.
Duarouter computes vehicle routes with an adjustable degree of variability, controlled by a parameter $w \in [1, +\infty)$ that controls the deviation of the generated route from the fastest route.
When $w$ = 1, duarouter calculates the standard fastest route. 
When $w>1$, it dynamically alters the edge weights (expected travel time) by a random factor uniformly drawn from the interval $[1,w)$. 
This random process ensures that different vehicles may get different routes even if their trip has the same origin and destination. 
As $w$ increases, the extent of randomness in the route calculation also increases, resulting in routes that can diverge substantially from the fastest route.
The injection of route variability helps us model the imperfections of human driving behaviour, which often deviates from the fastest route due to personal preferences, lack of complete knowledge of the road network, and irrational behaviours \cite{seele2012cognitive, zhu2015do}.
In our experiments, we perform simulations introducing a moderate level of randomness in the control group by varying $w \in \{3,5,7\}$. Our results are consistent across different values of $w$. For clarity and conciseness, we report in the main manuscript only the results for $w=5$; results for $w=3$ and $w=7$ are discussed in Supplementary Note 7.

\subsection{GPS data}
To estimate each city's mobility demand, we use a GPS dataset provided by OCTO that describes the trajectories of thousands of vehicles for various Italian localities over an entire year.
The market penetration of the dataset varies between 2-5\% of the total registered vehicles.
Our analysis focuses on Florence, Milan, and Rome (Italy). We selected these cities due to their diverse sizes, populations, and road network structures.
The raw dataset includes 7,102,351 GPS points from 24,640 vehicles in Florence, 143,698,720 GPS points from 106,456 vehicles in Milan, and 26,801,872 GPS points from 30,763 vehicles in Rome.
The dataset's validity and reliability are corroborated by its use in prior studies \cite{bazzani2011towards, pappalardo2015returners, Bohm2022-ck}.

We process the GPS dataset to create segmented trajectories that represent semantic journeys as follows: first, we filter out GPS points with speeds exceeding 250 km/h \cite{zheng2015trajectory}. 
Second, we use a stop detection algorithm \cite{zheng2015trajectory, pappalardo2022scikit} to segment each trajectory into sub-trajectories based on identified stops. 
A stop is identified when a vehicle remains within a distance of $0.2$ km from a trajectory point for at least 20 minutes. Third, we consider only trips that start and end within the predefined area of interest. 
For trips that start or end outside this area but traverse it, we used the first entry point within the area as the origin and the last exit point as the destination.
This method preserves commuter trips originating or ending outside the city.
Fourth, we keep all trips with a duration between 5 and 60 minutes that depart during the morning peak hours on each Wednesday.
This allows us to model high-traffic scenarios (as peak hours represent periods of intense traffic) and to capture typical weekday traffic, avoiding anomalies associated with weekends.
Finally, we exclude outlier holiday weeks from the analysis.
After the pre-processing step, the dataset includes 5,477 trips from 1,184 vehicles for Florence (616 distinct flows); 113,323 trips from 15,997 vehicles for Milan (11,967 distinct flows); and 13,647 trips from 1,965 vehicles for Rome (1,757 distinct flows).   
See Supplementary Note 1 for further details on the pre-processing.

\newpage

\section*{Data Availability}
The data generated in this study (i.e., simulation outputs including 
route diversity, \ce{CO_2} emissions, and other traffic measures, 
for all cities, navigation services, adoption rates, and traffic loads) 
are available at the repository 
\href{https://github.com/GiulianoCornacchia/Urban-Impact-Navigators}
{https://github.com/GiulianoCornacchia/Urban-Impact-Navigators} and permanently archived \cite{cornacchia2026reponavigators}.
The processed road network data used in this study are also available 
at the same repository.
The raw GPS trajectory data used in this study are not publicly available 
due to commercial restrictions; they were provided by OCTO under a 
data-sharing agreement and cannot be redistributed. To overcome this 
limitation, we have provided code to generate an OD matrix for Milan 
using a publicly accessible dataset available at 
\url{https://sobigdata.d4science.org/catalogue-sobigdata?path=/dataset/gps_track_milan_italy}, as well as a routine to create random OD matrices for scenarios where 
trajectory data is unavailable.
The navigation service route suggestions used in this study are not 
publicly available due to the terms of use of the respective APIs; 
we have instead provided code to generate the fastest route as a 
navigation service prototype, including routines to generate 
perturbations for non-routed vehicles, available at \href{https://github.com/GiulianoCornacchia/Urban-Impact-Navigators}
{https://github.com/GiulianoCornacchia/Urban-Impact-Navigators}.

\section*{Code Availability}
The Python code implementing the simulation framework and for replicating the analyses in the study is publicly available at \href{https://github.com/GiulianoCornacchia/Urban-Impact-Navigators}{https://github.com/GiulianoCornacchia/Urban-Impact-Navigators} and permanently archived at Zenodo \cite{cornacchia2026reponavigators}: 
\href{https://doi.org/10.5281/zenodo.20001710}
{https://doi.org/10.5281/zenodo.20001710}.

\section*{Funding}
G.C. and L.P. disclose support for the research of this work from 
the European Commission under the Next Generation EU programme, 
PNRR PRIN 2022 ``URBAI - Urban Artificial Intelligence'' 
[grant 20224CZ5X4 PE6, CUP B53D23012770006].
L.P. also discloses support for the research of this work from the Italian Project Fondo Italiano 
per la Scienza [grant FIS-2024-03129 CAIO, CUP B53C25003940001].
M.N. and D.P. disclose support for the research of this work from 
the European Commission under the Next Generation EU programme, 
PNRR - M4C2 - Investimento 1.3, Partenariato Esteso 
``FAIR - Future Artificial Intelligence Research'' - Spoke 1 
``Human-centered AI'' [grant PE00000013].

\section*{Acknowledgements} 
We thank Daniele Fadda for the invaluable support with the visualisations. 
We thank Gabriele Barlacchi, Matteo Böhm, Daniele Gambetta, Margherita Lalli, and Giovanni Mauro for the useful suggestions.

\section*{Author Contributions Statement} 
GC conceptualised the work, designed and developed the simulation framework, performed all the experiments, designed and made plots and figures, and wrote the paper. 
MN conceptualised the work. 
DP conceptualised the work and wrote the paper.
LP conceptualised the work, designed the experiments, designed plots and figures, wrote the paper, and supervised the research.
All authors read and approved the paper.

\section*{Competing Interests Statement} The authors declare no competing interests.

\bibliographystyle{naturemag}
\bibliography{biblio}

@article{ahn2013eco,
	title        = {Network-wide impacts of eco-routing strategies: A large-scale case study},
	author       = {Kyoungho Ahn and Hesham A. Rakha},
	year         = {2013},
	journal      = {Transportation Research Part D: Transport and Environment},
	volume       = {25},
	pages        = {119--130},
	doi          = {https://doi.org/10.1016/j.trd.2013.09.006},
	issn         = {1361-9209}
}

@article{argota2022getting,
	title        = {Getting Real: The Challenge of Building and Validating a Large-Scale Digital Twin of {Barcelona}'s Traffic with Empirical Data},
	author       = {Argota Sánchez-Vaquerizo, Javier},
	year         = {2022},
	journal      = {ISPRS International Journal of Geo-Information},
	volume       = {11},
	number       = {1},
	doi          = {10.3390/ijgi11010024},
	issn         = {2220-9964},
	article-number = {24}
}

@article{aronow2017estimating,
	title        = {Estimating Average Causal Effects Under General Interference, With Application to a Social Network Experiment},
	author       = {Aronow, Peter M. and Samii, Cyrus},
	year         = {2017},
	journal      = {The Annals of Applied Statistics},
	volume       = {11},
	number       = {4},
	pages        = {1912--1947}
}

@article{arora2021quantifying,
	title        = {Quantifying the sustainability impact of {Google Maps}: A case study of {Salt Lake City}},
	author       = {Arora, Neha and Cabannes, Theophile and Ganapathy, Sanjay and Li, Yechen and McAfee, Preston and Nunkesser, Marc and Osorio, Carolina and Tomkins, Andrew and Tsogsuren, Iveel},
	year         = {2021},
	journal      = {arXiv preprint arXiv:2111.03426}
}

@inproceedings{bachechi2020ModenaSensors,
	title        = {Using Real Sensors Data to Calibrate a Traffic Model for the City of {M}odena},
	author       = {Bachechi, Chiara and Rollo, Federica and Desimoni, Federico and Po, Laura},
	year         = {2020},
	booktitle    = {IHSI 2020},
	publisher    = {Springer},
	series       = {Advances in Intelligent Systems and Computing},
	volume       = {1131},
	pages        = {468--473},
	doi          = {10.1007/978-3-030-39512-4_73}
}

@inproceedings{barth2007environmental,
	title        = {Environmentally-Friendly Navigation},
	author       = {Barth, Matthew and Boriboonsomsin, Kanok and Vu, Alex},
	year         = {2007},
	booktitle    = {2007 IEEE Intelligent Transportation Systems Conference},
	volume       = {},
	number       = {},
	pages        = {684--689},
	doi          = {10.1109/ITSC.2007.4357672}
}

@inproceedings{bazzani2011towards,
	title        = {Towards congestion detection in transportation networks using GPS data},
	author       = {Bazzani, Armando and Giorgini, Bruno and Gallotti, Riccardo and Giovannini, Luca and Marchioni, Monica and Rambaldi, Sandro},
	year         = {2011},
	booktitle    = {2011 IEEE Third International Conference on Privacy, Security, Risk and Trust and 2011 IEEE Third International Conference on Social Computing},
	pages        = {1455--1459},
	organization = {IEEE}
}

@inproceedings{behrisch2011SUMOOverview,
	title        = {{SUMO}--simulation of urban mobility: an overview},
	author       = {Behrisch, Michael and Bieker, Laura and Erdmann, Jakob and Krajzewicz, Daniel},
	year         = {2011},
	booktitle    = {Proceedings of SIMUL 2011, The Third International Conference on Advances in System Simulation},
	organization = {ThinkMind}
}

@article{Bohm2022-ck,
	title        = {Gross polluters and vehicle emissions reduction},
	author       = {B{\"o}hm, Matteo and Nanni, Mirco and Pappalardo, Luca},
	year         = {2022},
	month        = jun,
	journal      = {Nature Sustainability},
    volume       = {5},
	publisher    = {Nature Publishing Group},
	pages        = {699--707},
	language     = {en}
}

@article{bokanyi2020understanding,
	title        = {{Understanding Inequalities in Ride-Hailing Services Through Simulations}},
	author       = {Bok{\'{a}}nyi, Eszter and Hann{\'{a}}k, Anik{\'{o}}},
	year         = {2020},
	journal      = {Scientific Reports},
	volume       = {10},
	number       = {1},
	pages        = {6500},
	issn         = {2045-2322}
}

@article{bonnel2018origin,
	title        = {Origin-Destination estimation using mobile network probe data},
	author       = {Bonnel, Patrick and Fekih, Mariem and Smoreda, Zbigniew},
	year         = {2018},
	journal      = {Transportation Research Procedia},
	publisher    = {Elsevier},
	volume       = {32},
	pages        = {69--81}
}

@article{cerqueira2022inference,
	title        = {Inference of dynamic origin--destination matrices with trip and transfer status from individual smart card data},
	author       = {Cerqueira, Sofia and Arsenio, Elisabete and Henriques, Rui},
	year         = {2022},
	journal      = {European Transport Research Review},
	publisher    = {SpringerOpen},
	volume       = {14},
	number       = {1},
	pages        = {1--18}
}

@inproceedings{cornacchia2022how,
	title        = {How Routing Strategies Impact Urban Emissions},
	author       = {Cornacchia, Giuliano and B{\"o}hm, Matteo and Mauro, Giovanni and Nanni, Mirco and Pedreschi, Dino and Pappalardo, Luca},
	year         = {2022},
	booktitle    = {Proceedings of the 30th International Conference on Advances in Geographic Information Systems},
	pages        = {1--4}
}

@software{cornacchia2026reponavigators,
	title        = {The traffic concentration effects of urban navigation services},
	author       = {Cornacchia, Giuliano and Nanni, Mirco and Pedreschi, Dino and Pappalardo, Luca},
	year         = {2026},
	month        = may,
	publisher    = {Zenodo},
	doi          = {10.5281/zenodo.20001710},
	url          = {https://doi.org/10.5281/zenodo.20001710},
	note         = {GitHub repository: \url{https://github.com/GiulianoCornacchia/Urban-Impact-Navigators}},
	version      = {v1.0.0}
}

@book{cox1958planning,
	title        = {Planning of experiments.},
	author       = {Cox, David Roxbee},
	year         = {1958},
	publisher    = {Wiley}
}

@article{di2016boundedly,
	title        = {Boundedly rational route choice behavior: A review of models and methodologies},
	author       = {Di, Xuan and Liu, Henry X},
	year         = {2016},
	journal      = {Transportation Research Part B: Methodological},
	publisher    = {Elsevier},
	volume       = {85},
	pages        = {142--179}
}

@inproceedings{ensign2018runaway,
	title        = {Runaway feedback loops in predictive policing},
	author       = {Ensign, Danielle and Friedler, Sorelle A and Neville, Scott and Scheidegger, Carlos and Venkatasubramanian, Suresh},
	year         = {2018},
	booktitle    = {Conference on fairness, accountability and transparency},
	pages        = {160--171},
	organization = {PMLR}
}

@misc{EU2022DSA,
	title        = {{Regulation (EU) 2022/2065 of the European Parliament and of the Council of 19 October 2022 on a Single Market for Digital Services and amending Directive 2000/31/EC (Digital Services Act)}},
	year         = {2022},
	note         = {Accessed via EUR-Lex},
	howpublished = {\url{https://eur-lex.europa.eu/legal-content/EN/TXT/?uri=CELEX:32022R2065}},
    author = {{European Parliament and Council}},
	institution  = {European Parliament and Council}
}

@inproceedings{flitsch2018UpperAustriaSUMO,
	title        = {Calibrating Traffic Simulation Models in {SUMO} Based upon Diverse Historical Real--Time Traffic Data -- Lessons Learned in ITS Upper {Austria}},
	author       = {Flitsch, Christina and Kastner, Karl-Heinz and B{\'o}sa, K{\'a}roly and Neubauer, Matthias},
	year         = {2018},
	booktitle    = {SUMO2018 (EPiC Series in Engineering)},
	volume       = {2},
	pages        = {25--42}
}

@article{foderaro2017navigation,
	title        = {Navigation apps are turning quiet neighborhoods into traffic nightmares},
	author       = {Foderaro, Lisa W.},
	year         = {2017},
	month        = {Dec 24},
	journal      = {The New York Times},
	url          = {https://www.nytimes.com/2017/12/24/nyregion/traffic-apps-gps-neighborhoods.html},
	note         = {Accessed: 2025-01-12}
}

@book{harvey1996justice,
	title        = {Justice, nature and the geography of difference},
	author       = {Harvey, David},
	year         = {1996},
	publisher    = {Blackwell Publishers}
}

@article{hazelton2001inference,
	title        = {Inference for origin--destination matrices: estimation, prediction and reconstruction},
	author       = {Hazelton, Martin L},
	year         = {2001},
	journal      = {Transportation Research Part B: Methodological},
	publisher    = {Elsevier},
	volume       = {35},
	number       = {7},
	pages        = {667--676}
}

@article{hendrix2016traffic,
	title        = {Traffic-weary homeowners and Waze are at war, again. Guess who’s winning?},
	author       = {Hendrix, Steve},
	year         = {2016},
	month        = {Jun 5},
	journal      = {The Washington Post},
	url          = {https://www.washingtonpost.com/local/traffic-weary-homeowners-and-waze-are-at-war-again-guess-whos-winning/2016/06/05/c466df46-299d-11e6-b989-4e5479715b54_story.html},
	note         = {Accessed: 2025-01-12}
}

@article{huszar2022algorithmic,
	title        = {Algorithmic amplification of politics on {Twitter}},
	author       = {Husz{\'a}r, Ferenc and Ktena, Sofia Ira and O'Brien, Conor and Belli, Luca and Schlaikjer, Andrew and Hardt, Moritz},
	year         = {2022},
	month        = jan,
	journal      = {Proceedings of the National Academy of Sciences},
	volume       = {119},
    pages   = {e2025334119},
	number       = {1}
}

@misc{infras2013handbook,
	title        = {Handbuch für Emissionsfaktoren},
	author       = {INFRAS},
	year         = {2013},
	howpublished = {\url{http://www.hbefa.net/}}
}

@article{johnson2017beautiful,
	title        = {Beautiful… but at what cost? An examination of externalities in geographic vehicle routing},
	author       = {Johnson, Isaac and Henderson, Jessica and Perry, Caitlin and Sch{\"o}ning, Johannes and Hecht, Brent},
	year         = {2017},
	journal      = {Proceedings of the ACM on Interactive, Mobile, Wearable and Ubiquitous Technologies},
	publisher    = {ACM New York, NY, USA},
	volume       = {1},
	number       = {2},
	pages        = {1--21},
	doi          = {10.1145/3090080}
}

@inproceedings{krajzewicz2002sumo,
	title        = {{SUMO} (Simulation of Urban MObility)-an open-source traffic simulation},
	author       = {Krajzewicz, Daniel and Hertkorn, Georg and R{\"o}ssel, Christian and Wagner, Peter},
	year         = {2002},
	booktitle    = {Proceedings of the 4th middle East Symposium on Simulation and Modelling (MESM20002)},
	pages        = {183--187}
}

@inproceedings{krajzewicz2015second,
	title        = {Second Generation of Pollutant Emission Models for {SUMO}},
	author       = {Krajzewicz, Daniel and Behrisch, Michael and Wagner, Peter and Luz, Raphael and Krumnow, Mario},
	year         = {2015},
	booktitle    = {Modeling Mobility with Open Data},
	publisher    = {Springer International Publishing},
	address      = {Cham},
	pages        = {203--221},
	editor       = {Behrisch, Michael and Weber, Melanie}
}

@article{lee2019recommender,
	title        = {How do recommender systems affect sales diversity? A cross-category investigation via randomized field experiment},
	author       = {Lee, Dokyun and Hosanagar, Kartik},
	year         = {2019},
	journal      = {Information Systems Research},
	publisher    = {INFORMS},
	volume       = {30},
	number       = {1},
	pages        = {239--259}
}

@article{mauro2026urban,
	title        = {The urban impact of {AI}: modelling feedback loops in location-based recommender systems},
	author       = {Mauro, Giovanni and Minici, Marco and Pappalardo, Luca},
	year         = {2026},
	month        = jan,
	journal      = {Machine Learning},
	volume       = {115},
	number       = {1},
	pages        = {19},
	doi          = {10.1007/s10994-025-06904-z},
	issn         = {1573-0565}
}

@inproceedings{microscopic2018lopez,
	title        = {Microscopic Traffic Simulation using {SUMO}},
	author       = {Lopez, Pablo Alvarez and Behrisch, Michael and Bieker-Walz, Laura and Erdmann, Jakob and Flötteröd, Yun-Pang and Hilbrich, Robert and Lücken, Leonhard and Rummel, Johannes and Wagner, Peter and Wiessner, Evamarie},
	year         = {2018},
	booktitle    = {2018 21st International Conference on Intelligent Transportation Systems (ITSC)},
	volume       = {},
	number       = {},
	pages        = {2575--2582},
	doi          = {10.1109/ITSC.2018.8569938}
}

@article{ngo2021effects,
	title        = {The effects of ride-hailing services on bus ridership in a medium-sized urban area using micro-level data: Evidence from the Lane Transit District},
	author       = {Ngo, Nicole S and G{\"o}tschi, Thomas and Clark, Benjamin Y},
	year         = {2021},
	journal      = {Transport Policy},
	publisher    = {Elsevier},
	volume       = {105},
	pages        = {44--53}
}

@article{pappalardo2015returners,
	title        = {Returners and explorers dichotomy in human mobility},
	author       = {Pappalardo, Luca and Simini, Filippo and Rinzivillo, Salvatore and Pedreschi, Dino and Giannotti, Fosca and Barab{\'a}si, Albert-L{\'a}szl{\'o}},
	year         = {2015},
	month        = sep,
	journal      = {Nature Communications},
	volume       = {6},
	pages        = {8166},
	language     = {en}
}

@article{pappalardo2022scikit,
	title        = {scikit-mobility: A Python Library for the Analysis, Generation, and Risk Assessment of Mobility Data},
	author       = {Pappalardo, Luca and Simini, Filippo and Barlacchi, Gianni and Pellungrini, Roberto},
	year         = {2022},
	journal      = {Journal of Statistical Software},
	volume       = {103},
	number       = {4},
	pages        = {1--38},
	doi          = {10.18637/jss.v103.i04}
}

@article{pappalardo2023future,
	title        = {{Future directions in human mobility science}},
	author       = {Pappalardo, Luca and Manley, Ed and Sekara, Vedran and Alessandretti, Laura},
	year         = {2023},
	journal      = {Nature Computational Science},
	volume       = {3},
	pages        = {588--600},
	doi          = {10.1038/s43588-023-00469-4},
	issn         = {2662-8457}
}

@article{pappalardo2025survey,
	title        = {A survey on the impacts of recommender systems on users, items, and human-{AI} ecosystems},
	author       = {Pappalardo, Luca and Citraro, Salvatore and Cornacchia, Giuliano and Nanni, Mirco and Pansanella, Valentina and Rossetti, Giulio and Gezici, Gizem and Giannotti, Fosca and Lalli, Margherita and Mauro, Giovanni and others},
	year         = {2025},
	journal      = {arXiv preprint arXiv:2407.01630}
}

@article{pedreschi2025human,
	title        = {Human-{AI} coevolution},
	author       = {Pedreschi, Dino and Pappalardo, Luca and Ferragina, Emanuele and Baeza-Yates, Ricardo and Barab{\'a}si, Albert-L{\'a}szl{\'o} and Dignum, Frank and Dignum, Virginia and Eliassi-Rad, Tina and Giannotti, Fosca and Kert{\'e}sz, J{\'a}nos and others},
	year         = {2025},
	journal      = {Artificial Intelligence},
	publisher    = {Elsevier},
	volume       = {339},
	pages        = {104244},
	doi          = {10.1016/j.artint.2024.104244}
}

@article{perezprada2017managing,
	title        = {Managing Traffic Flows for Cleaner Cities: The Role of Green Navigation Systems},
	author       = {Perez-Prada, Fiamma and Monzón, Andrés and Valdés, Cristina},
	year         = {2017},
	month        = {06},
	journal      = {Energies},
	volume       = {10},
	pages        = {791},
	doi          = {10.3390/en10060791}
}

@article{rahwan2019machine,
	title        = {Machine behaviour},
	author       = {Rahwan, Iyad and Cebrian, Manuel and Obradovich, Nick and Bongard, Josh and Bonnefon, Jean-Fran{\c{c}}ois and Breazeal, Cynthia and Crandall, Jacob W and Christakis, Nicholas A and Couzin, Iain D and Jackson, Matthew O and others},
	year         = {2019},
	journal      = {Nature},
	publisher    = {Nature Publishing Group UK London},
	volume       = {568},
	number       = {7753},
	pages        = {477--486}
}

@article{sanchez2023bias,
	title        = {Bias characterization, assessment, and mitigation in location-based recommender systems},
	author       = {S{\'a}nchez, Pablo and Bellog{\'\i}n, Alejandro and Boratto, Ludovico},
	year         = {2023},
	journal      = {Data Mining and Knowledge Discovery},
	publisher    = {Springer},
	volume       = {37},
	number       = {5},
	pages        = {1885--1929}
}

@article{santi2014quantifying,
	title        = {Quantifying the benefits of vehicle pooling with shareability networks},
	author       = {Santi, Paolo and Resta, Giovanni and Szell, Michael and Sobolevsky, Stanislav and Strogatz, Steven H and Ratti, Carlo},
	year         = {2014},
	journal      = {Proceedings of the National Academy of Sciences},
	publisher    = {National Academy of Sciences},
	volume       = {111},
	number       = {37},
	pages        = {13290--13294}
}

@article{schade2024traffic,
	title        = {Traffic jam by {GPS}: A systematic analysis of the negative social externalities of large-scale navigation technologies},
	author       = {Schade, Eve AND Savino, Gian-Luca AND Gunal, Yasemin AND Schöning, Johannes},
	year         = {2024},
	month        = {08},
	journal      = {PLOS ONE},
	publisher    = {Public Library of Science},
	volume       = {19},
	number       = {8},
	pages        = {1--18},
	doi          = {10.1371/journal.pone.0308260},
	url          = {https://doi.org/10.1371/journal.pone.0308260}
}

@book{schlosberg2007defining,
	title        = {Defining environmental justice: Theories, movements, and nature},
	author       = {Schlosberg, David},
	year         = {2007},
	publisher    = {OUP Oxford}
}

@article{seele2012cognitive,
	title        = {Cognitive aspects of traffic simulations in virtual environments},
	author       = {Seele, Sven and Dettmar, Thomas and Herpers, Rainer and Bauckhage, Christian and Becker, Peter},
	year         = {2012},
	journal      = {SNE Simul. Notes Eur.},
	publisher    = {ARGESIM Arbeitsgemeinschaft Simulation News},
	volume       = {22},
	number       = {2},
    pages   =    {83--88},
}

@book{sheffi1984urban,
	title        = {Urban Transportation Networks: Equilibrium Analysis with Mathematical Programming Methods},
	author       = {Sheffi, Yosef},
	year         = {1984},
	publisher    = {Prentice-Hall},
	address      = {Englewood Cliffs, NJ, USA}
}

@article{shumailov2024ai,
	title        = {AI models collapse when trained on recursively generated data},
	author       = {Shumailov, Ilia and Shumaylov, Zakhar and Zhao, Yiren and Papernot, Nicolas and Anderson, Ross and Gal, Yarin},
	year         = {2024},
	journal      = {Nature},
	publisher    = {Nature},
	volume       = {631},
	number       = {8022},
	pages        = {755--759}
}

@article{Siuhi2016-gb,
	title        = {Opportunities and challenges of smart mobile applications in transportation},
	author       = {Siuhi, Saidi and Mwakalonge, Judith},
	year         = {2016},
	journal      = {Journal of Traffic and Transportation Engineering},
	volume       = {3},
	number       = {6},
	pages        = {582--592}
}

@inproceedings{thai2016negative,
	title        = {Negative externalities of {GPS}-enabled routing applications: A game theoretical approach},
	author       = {Thai, Jérôme and Laurent-Brouty, Nicolas and Bayen, Alexandre M.},
	year         = {2016},
	booktitle    = {2016 IEEE 19th International Conference on Intelligent Transportation Systems (ITSC)},
	volume       = {},
	number       = {},
	pages        = {595--601},
	doi          = {10.1109/ITSC.2016.7795614}
}

@inproceedings{valdes2016eco,
	title        = {Eco-routing: More green drivers means more benefits},
	author       = {Valdes, C and Perez-Prada, F and Monzon, A},
	year         = {2016},
	booktitle    = {Proceedings of the XII Conference on Transport Engineering, Valencia, Spain},
	pages        = {7--9}
}

@article{wardrop1952proceedings,
  title   = {Some theoretical aspects of road traffic research},
  author  = {Wardrop, J. G.},
  year    = {1952},
  journal = {Proceedings of the Institution of Civil Engineers},
  volume  = {1},
  number  = {3},
  pages   = {325--362}
}

@inproceedings{yan2020fairness,
	title        = {{Fairness-Aware} Demand Prediction for New Mobility},
	author       = {Yan, An and Howe, Bill},
	year         = {2020},
	booktitle    = {Proceedings of the {AAAI} Conference on Artificial Intelligence},
	volume       = {34},
	pages        = {1079--1087}
}

@article{zheng2015trajectory,
	title        = {Trajectory Data Mining: An Overview},
	author       = {Zheng, Yu},
	year         = {2015},
	month        = {may},
	journal      = {ACM Trans. Intell. Syst. Technol.},
	publisher    = {Association for Computing Machinery},
	address      = {New York, NY, USA},
	volume       = {6},
	number       = {3},
	doi          = {10.1145/2743025},
	issn         = {2157-6904},
	issue_date   = {May 2015},
	articleno    = {29},
	numpages     = {41}
}

@article{zhu2015do,
	title        = {Do People Use the Shortest Path? An Empirical Test of Wardrop’s First Principle},
	author       = {Zhu, Shanjiang AND Levinson, David},
	year         = {2015},
	month        = {08},
	journal      = {PLOS ONE},
	publisher    = {Public Library of Science},
	volume       = {10},
	number       = {8},
	pages        = {1--18},
	doi          = {10.1371/journal.pone.0134322}
}

\end{document}